\documentclass[sigchi]{acmart}

\AtBeginDocument{%
  \providecommand\BibTeX{{%
    \normalfont B\kern-0.5em{\scshape i\kern-0.25em b}\kern-0.8em\TeX}}}

\setcopyright{rightsretained} 
\begin{document}

\title[Poster: Efficient Convolutional Neural Network for FMCW Radar ...]{Poster: Efficient Convolutional Neural Network for FMCW Radar Based Hand Gesture Recognition}

\copyrightyear{2019} \acmYear{2019} 
\acmConference[UbiComp/ISWC '19 Adjunct]{Adjunct Proceedings of the 2019 ACM International Joint Conference on Pervasive and Ubiquitous Computing and the 2019 International Symposium on Wearable Computers}{September 9--13, 2019}{London, United Kingdom} 
\acmBooktitle{Adjunct Proceedings of the 2019 ACM International Joint Conference on Pervasive and Ubiquitous Computing and the 2019 International Symposium on Wearable Computers (UbiComp/ISWC '19 Adjunct), September 9--13, 2019, London, United Kingdom}\acmDOI{10.1145/3341162.3343768} \acmISBN{978-1-4503-6869-8/19/09} 

\author{Xiaodong Cai}
\email{xiaodong.cai@intel.com}
\affiliation{%
  \institution{Intel Corporation}
  \city{Shanghai}
  \country{China}
}

\author{Jingyi Ma}
\email{jingyi.ma@intel.com}
\affiliation{%
  \institution{Intel Labs China}
  \city{Beijing}
  \country{China}
}
\author{Wei Liu}
\email{wei.liu@intel.com}
\affiliation{%
  \institution{Intel Labs China}
  \city{Beijing}
  \country{China}
}
\author{Hemin Han}
\email{hemin.hani@intel.com}
\affiliation{%
  \institution{Intel Corporation}
  \city{Shanghai}
  \country{China}
}
\author{Lili Ma}
\email{michael.ma@intel.com}
\affiliation{%
  \institution{Intel Corporation}
  \city{Shanghai}
  \country{China}
}

\renewcommand{\shortauthors}{Cai et al.}

\begin{abstract}
FMCW radar could detect object's range, speed and Angle-of-Arrival, 
advantages are robust to bad weather, good range resolution, and good speed resolution. 
In this paper, we consider the FMCW radar as a novel interacting interface on laptop. 
We merge sequences of object's range, speed, azimuth information into single input,
then feed to a convolution neural network to learn spatial and temporal patterns.
Our model achieved 96\% accuracy on test set and real-time test.
\end{abstract}



\maketitle

\section{Introduction}
\begin{figure}
  \centering
  \includegraphics[width=1.\linewidth]{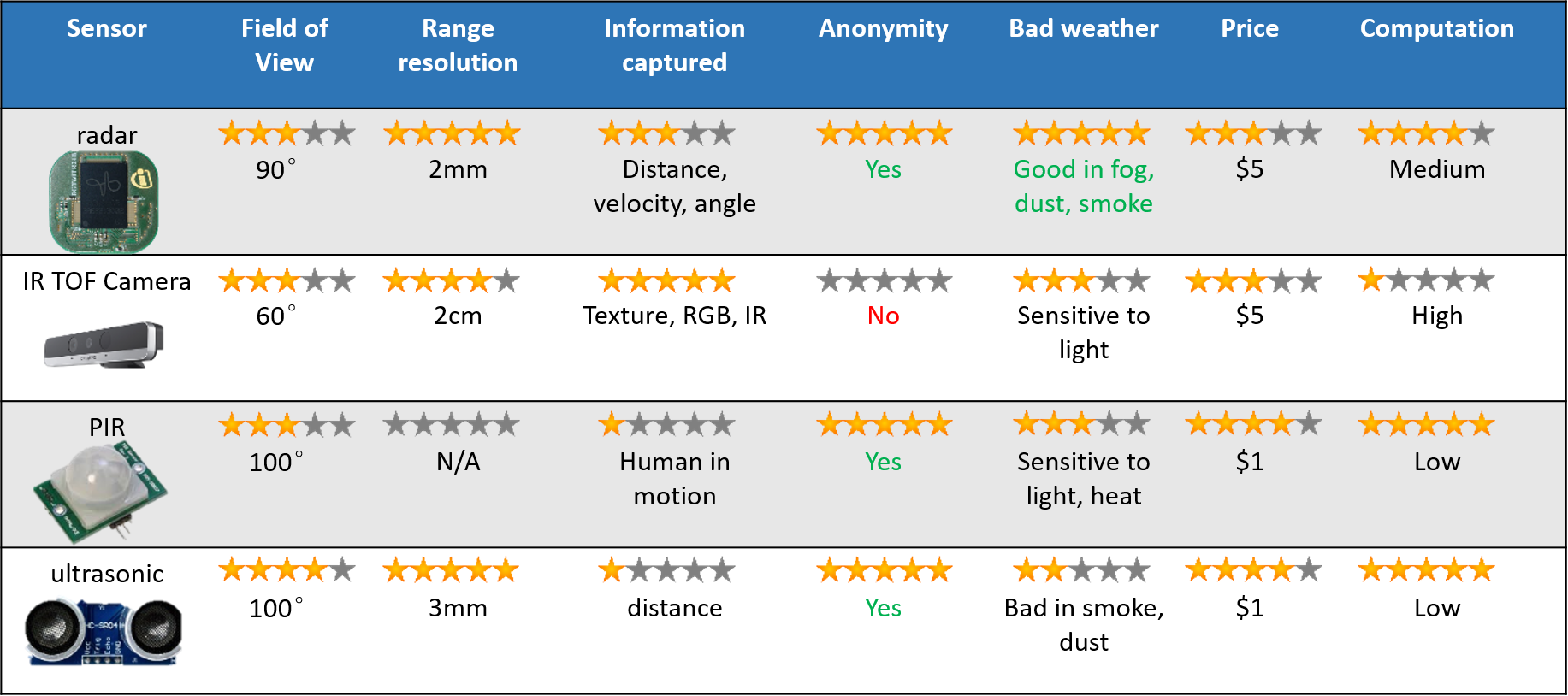}
  \caption{Sensors comparison}
  \label{sensors}
\end{figure}

For camera based gesture recognition, there are many commercial solutions, such as Kinect, leap motion, RealSense. These solutions suffer from privacy issues, while FMCW radar has no such limits. As  FMCW radar can only estimate object’s range, speed and angle information, so it capture human actions, but the information is not enough to identify the user. Detailed comparison among different sensors are shown in Figure \ref{sensors}. It is also friendly for industrial design that radar doesn’t need hole-punch, while microphone and camera does.

\subsection{FMCW Radar Basics}
\begin{figure}
  \centering
  \includegraphics[width=1.0\linewidth]{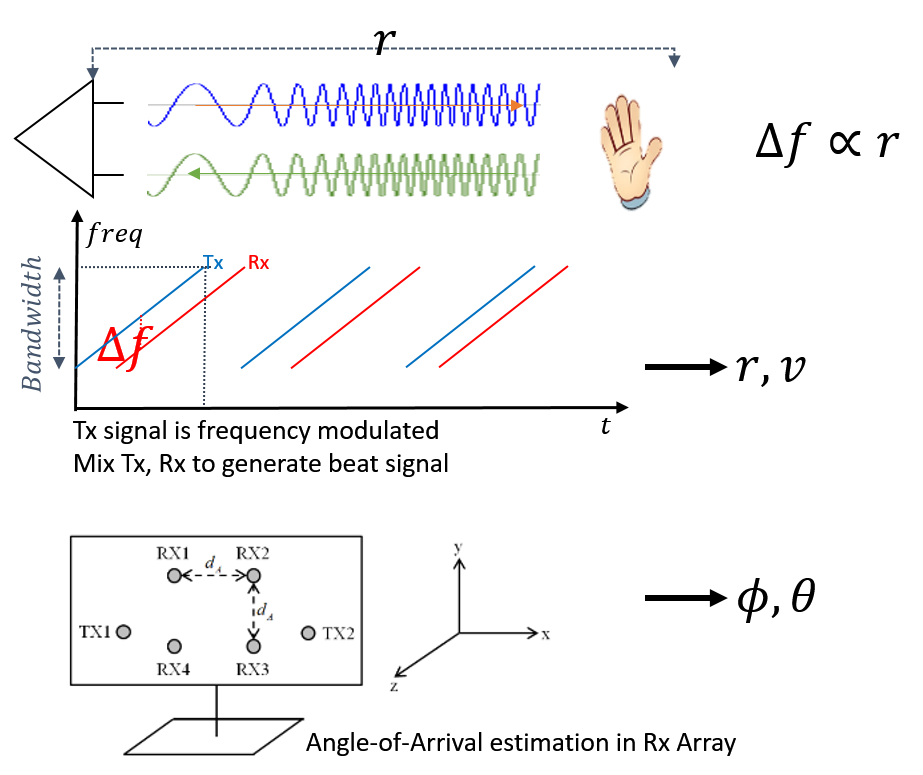}
  \caption{Radar basics}
  \label{radarBasic}
\end{figure}

\begin{figure}
  \centering
  \includegraphics[width=0.8\linewidth]{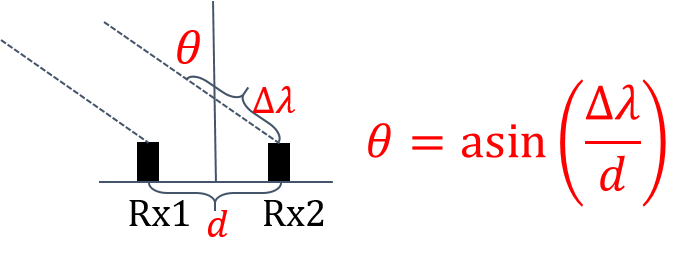}
  \caption{Angle of Arrival estimation}
  \label{aoa}
\end{figure}

FMCW radar is short for Frequency Modulated Continuous Wave Radar. 
Radar transmits a continuous carrier modulated by a periodic function such as a sinusoid wave to provide range data. 
At each period $T_c$ (also called a chirp), 
radar transmitter emits a sinusoidal wave, 
with frequency modulated from $f_{min}$ to $f_{max}$, 
Bandwidth $B=f_{max}-f_{min}$, which is proportional to radar spatial resolution. 
We set modulation slope $\alpha=B/T_c$.
In this paper, we use a 2Tx, 4Rx FMCW Radar, sweeping frequency 57-64GHz, 

Consider an object, e.g. palm of hand, 
with an initial range $R_0$ at $t_0 = 0$, 
and radial velocity $v$, where $v \ll c$, 
so the signal travel time $\tau=\frac{2}{c}(R_0+v\cdot t)$. 
FMCW radar transmitter emits sinusoidal signal, 
$s(t)=A\cdot cos(\phi_s \cdot t)$, 
the received signal $\phi_r(t)=s(t-\tau)$, 
mixed signal  $\phi_w(t)=\int(\phi_s(t)\phi_r^*(t)$, 
where the beat frequency in $\phi_w(t)$ is the range of the object; 
and the speed of the object can be estimated by frequency shift $f_d$, $v=\frac{f_d\lambda}{2}$.
Angle-of-Arrival is calculated via phase difference between 2 receivers. 
Same range bin and speed bin in Range-Doppler Map (RDM), 
Receiver 1 $r^{(i,j)}_1$ and Receiver 2 $r^{(i,j)}_2$,
 Angle-Of-Arrival $\theta = sin(\frac{\Delta \lambda}{d})$, 
where $\Delta \lambda =  \frac{2 \pi}{\lambda} \Delta \beta$, 
$\Delta \beta=atan(\frac{r^{(i,j)}_1.imag}{r^{(i,j)}_1.real}) - atan(\frac{r^{(i,j)}_2.imag}{r^{(i,j)}_2.real})$, illustrated in Figure \ref{aoa}.
How FMCW radar estimate hand movement is illustrated in Figure \ref{radarBasic}, and readers can refer to \cite{patole2017automotive} for technical details.

All in all, for a Multiple Input Multiple Output (MIMO) radar system, 
$Q$ objects' range, speed, angle can be estimated via 3D FFT, 
where  l is the antenna index,  
$n$ is sampling index,  
$p$ is the chirp index, 
$Q$ is the number of object, 
$\alpha^{(q)}$ is the amplitude factor,  
$K$ is modulation slope, 
$R^{(q)}$ is range of $q$ th  object, 
$f_d^{(q)}$   is the frequency shift of $q$ th  object, 
$f_c$ is radar’s center frequency, 
$d$ is the distance between antenna, 
$\theta^{(q)}$ is the azimuth of $q$ th object, 
$c$ is the speed of light.
$$ d(l,n,p)\approx \sum_{q=0}^{Q-1} \alpha^{(q)} \cdot e^{j2\pi [\frac{(\frac{2KR^{(q)}}{c} + f_d^{(q)})n}{f_s} + \frac{f_c l dsin\theta^{(q)}}{c} + f_d q p T_0 + \frac{2f_c R^{(q)}}{c}]} $$


\begin{figure}
  \centering
  \includegraphics[width=1.0\linewidth]{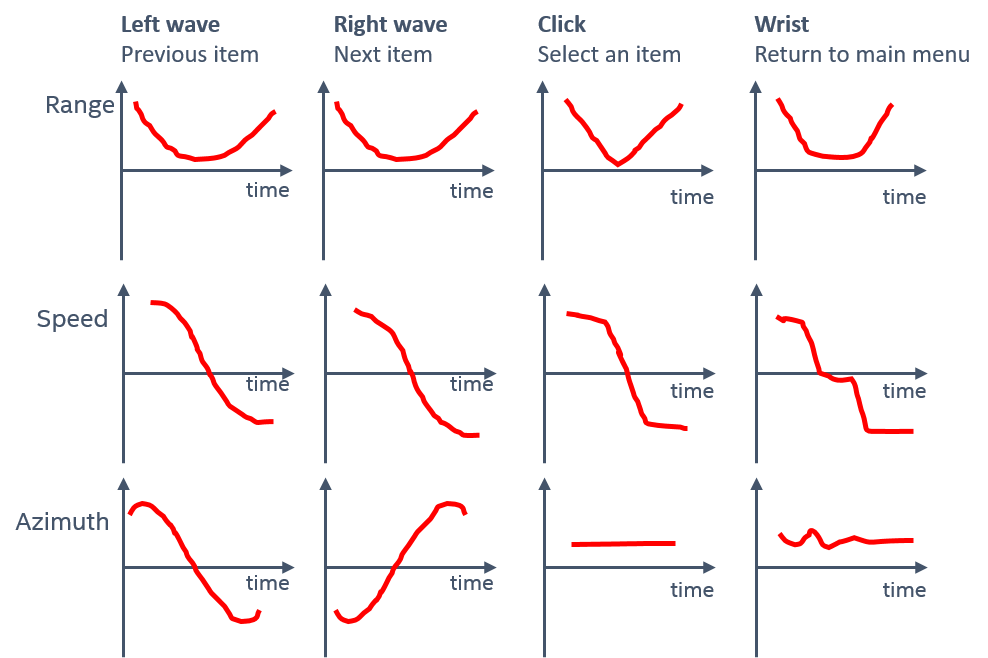}
  \caption{Theoretical analysis of hand movement}
  \Description{Theoretical analysis of hand movement}
  \label{gesture_analysis}
\end{figure}

\subsection{Gesture Definitions}

we defined four gestures: left wave, right wave, click, and wrist in Table \ref{tab:gesture_definition}.
We also plot theoretical analysis of range, speed, azimuth trajectories on the predefined gestures set, shown in Figure \ref{gesture_analysis}. From Figure \ref{gesture_analysis}, those gesture trajectories are quite distinctive. We build a template-matching algorithm for gesture recognition, the accuracy is ~70-80\%, and it is hard to extract trajectories from noisy radar signal.

\begin{table}
  \caption{Gesture Definitions}
  \label{tab:gesture_definition}
  \begin{tabular}{lll}
    \toprule
    Gesture&Hand Movement&Meaning of gesture\\
    \midrule
    LEFT & Move  from right to left & Browse previous item\\
    RIGHT & Move  from left to right & Browse next item\\
    CLICK & Finger pointing to radar & Select an item\\
    WRIST & Hand making fist & Return to main menu\\
  \bottomrule
\end{tabular}
\end{table}

\section{Related Work}

Previous work \cite{hazra2018robust},\cite{wang2016interacting}, \cite{zhang2018latern} are based on CNN+LSTM.
\cite{wang2016interacting} first introduced CNN+LSTM architecture to process radar based gesture recognition. CNN learns spatial patterns inside RDM, then feed into LSTM to learn temporal patterns among RDMs. CNN+LSTM achieved 87\% accuracy on 11 gestures.
\cite{hazra2018robust} replace CNN with AllConvNet to reduce parameters and inference time.
\cite{zhang2018latern} introduced 3D CNN to learn spatial and temporal patterns at the same time.

Solutions above \cite{hazra2018robust},\cite{wang2016interacting}, \cite{zhang2018latern} use real value Range-Doppler map as input, they ignore Angle-of-Arrival information. In order to take Angle-Of-Arrival into consideration, we need extra signal processing procedure.

\section{Proposed System}

\begin{figure}
  \centering
  \includegraphics[width=1.0\linewidth]{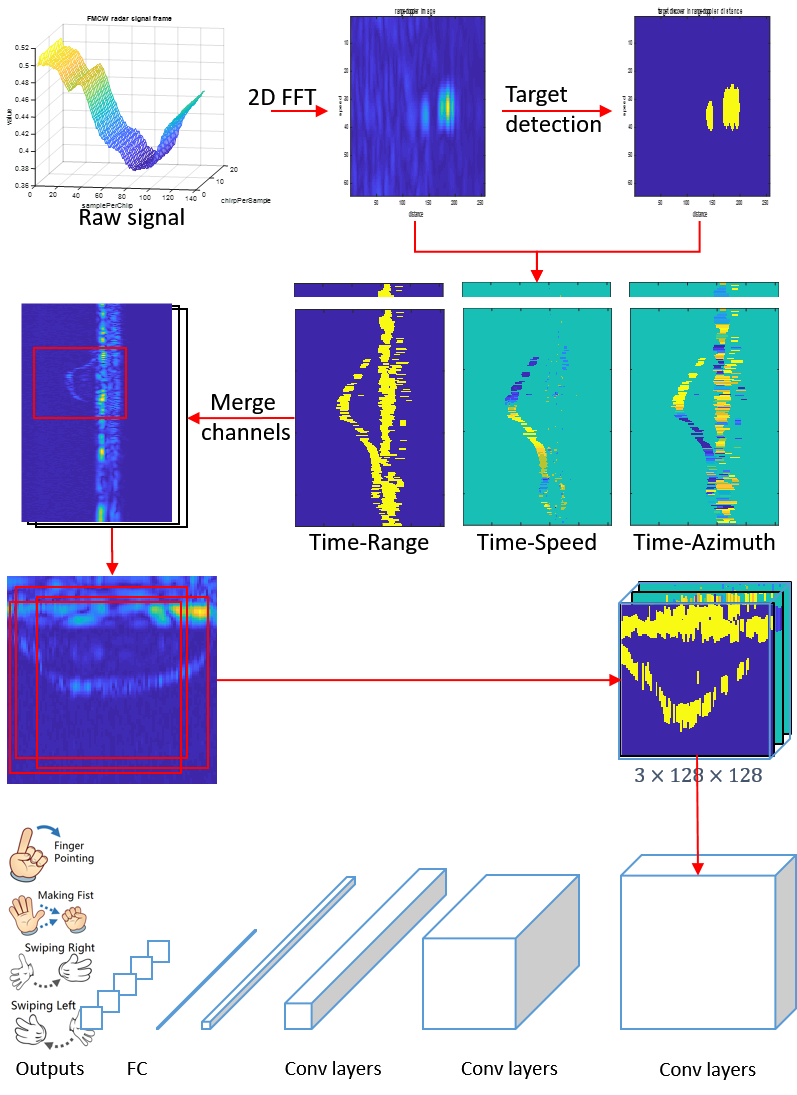}
  \caption{Radar data processing pipeline}
  \Description{radar data processing pipeline}
  \label{pipeline}
\end{figure}

To explicitly extract Range, Speed and Azimuth trajectory, 
we merge 128 RDM frames, each frame size is $64\times 256$, into a 3-channel input frame,
representing range-time, speed-time, azimuth-time respectively, RSA for short
The merged RSA input shape is $128\times128\times3$, then feed it into CNN.
Gesture recognition pipeline is shown in Table \ref{tab:signal_processing} and Figure \ref{pipeline}.

\begin{table*}
  \caption{Radar signal pre-Processing}
  \label{tab:signal_processing}
  \begin{tabular}{llll}
    \toprule
    Step & input shape & procedure & output shape\\
    \midrule
    \texttt 1 & $16\times 128$ & 2D FFT, to convert raw signal into RDM & $64\times 256$ \\
    \texttt 2 & $64\times 256$ & Do Constant False Alarm Rate (CFAR) \cite{blake1988cfar} on RDM to detect hand and body & $64\times 256$ \\
    \texttt 3 & $64\times 128$ & Crop RDM to keep body and hand, generate subset of RDM & $64\times 128$ \\
    \texttt 4 & $64\times 128$ & For each range bin, calculate maximum speed, average azimuth, generate a frame & $1\times 128 \times 3$ \\
    \texttt 5 & $1\times 128 \times 3$ & merge 128 frames of above output into one frame & $128\times128\times3$ \\
   
    \bottomrule
  \end{tabular}
\end{table*}

\subsection{Neural Network Architecture Desgin}
\begin{figure}
  \centering
  \includegraphics[width=1.0\linewidth]{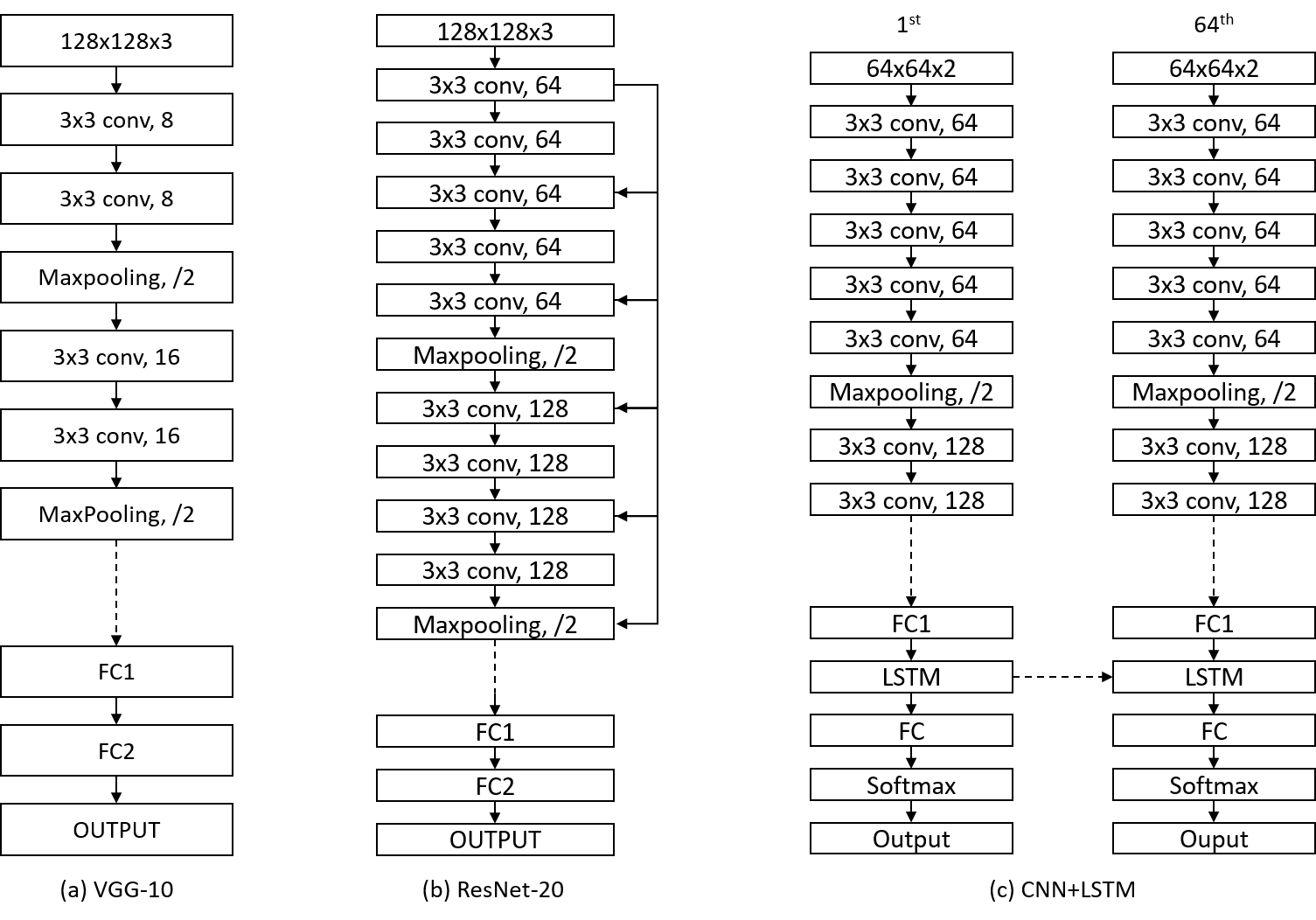}
  \caption{Network architecture, (a) VGG-10, (b) ResNet-20, (c) CNN+LSTM}
  \Description{Network Architecture}
  \label{network_arch}
\end{figure}

Firstly, we design a VGG-like neural network, 
 called VGG-10 in Figure \ref{network_arch}(a), 
follows \{Conv3x3, Conv3x3, MaxPooling\} building block, and 2 fully Connected layers.
 ADAM optimizer, early stopping and reduced learning rate is applied. 
VGG-10 converged at 10th epoch with validation accuracy 92\%.

To improve the performance, we add residual block between convolution layers, 
batch normalization is also added between each residual block to make back-propagation more robust \cite{he2016deep}, 
and we called it ResNet-20 in Figure \ref{network_arch}(b).
ResNet-20 outperforms VGG-10 achieved 98\% validation accuracy. 

We also build CNN+LSTM model for comparison, in Figure \ref{network_arch}(c). CNN+LSTM needs RDM sequence as input, first we resize original RDM ($64 \times 256$) to $64\times 64$, and feed 64 resized RDM frames into CNN. CNN module follows \{Conv5x5, Conv5x5, MaxPooling\} building block, and 1 fully Connected layer to encode feature, then put feature encoding into LSTM.

\section{Dataset and Experiment results}
We collected 50 subjects' gesture data, each subject did 4 gestures 10 times with left hand and right hand, total 3652 valid records, 
validation-train split ratio is 0.3.

\begin{figure}
  \centering
  \includegraphics[width=1.0\linewidth]{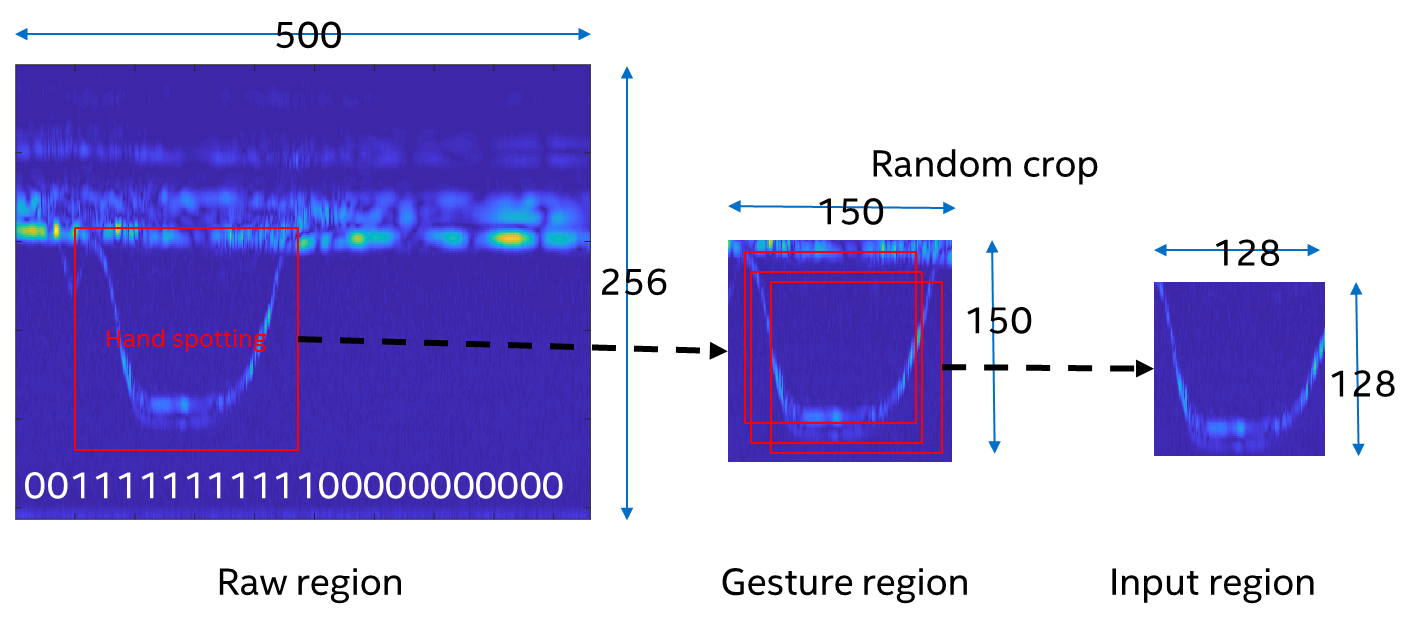}
  \caption{Data augmentation}
  \Description{data augmentation}
  \label{random_crop}
\end{figure}

We also do data-augmentation to enrich the dataset.
First, we draw a block containing gesture movements, 
then crop the area randomly to generate training data, shown in Figure \ref{random_crop}.
At last, we obtained more than 400k training data.

We compare VGG-10, ResNet-20 and CNN+LSTM on a same dataset, and calculate average accuracy. CNN+LSTM has the lowest accuracy on LEFT/RIGHT, due to lack of Angle-Of-Arrival information; Deep CNN outperforms shallow CNN, which is aligned with experiment results in \cite{wang2016interacting}.

\begin{table}
  \caption{Accuracy comparison among models}
  \label{tab:model_comparison}
  \begin{tabular}{llllll}
    \toprule
    \shortstack{Network \\ Architecture} & Avg. Acc. & LEFT & RIGHT & CLICK & WRIST\\
    \midrule
    \texttt VGG-10 & 91.0\% &  94.9\% &  80.7\% &  95.5\% &  97.0\% \\
    \texttt ResNet-20 & 98.7\% &  99.1\% &  99.0\% &  97.9\% &  98.9\% \\
    \texttt CNN+LSTM & 78.0\% &  69.0\% &  49.5\% &  84.6\% &  90.1\% \\
    \bottomrule
  \end{tabular}
\end{table}

\begin{figure}
  \centering
  \includegraphics[width=0.9\linewidth]{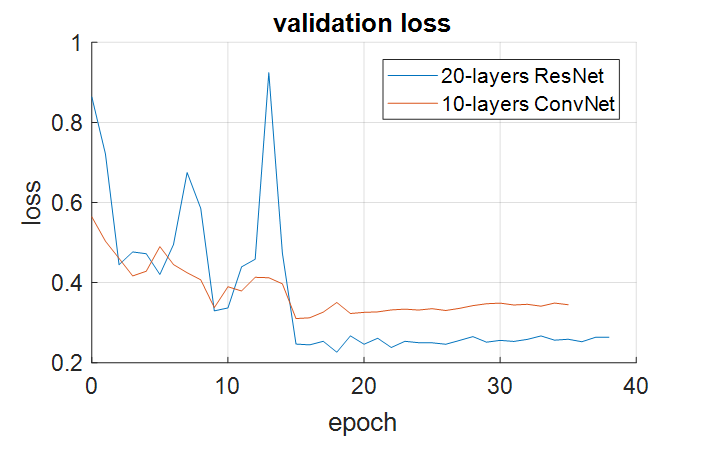}
  \caption{Validation loss of VGG-10 and ResNet-20}
  \Description{validation loss of VGG-10 and ResNet-20}
  \label{val_loss}
\end{figure}

\begin{figure}
  \centering
  \includegraphics[width=0.8\linewidth]{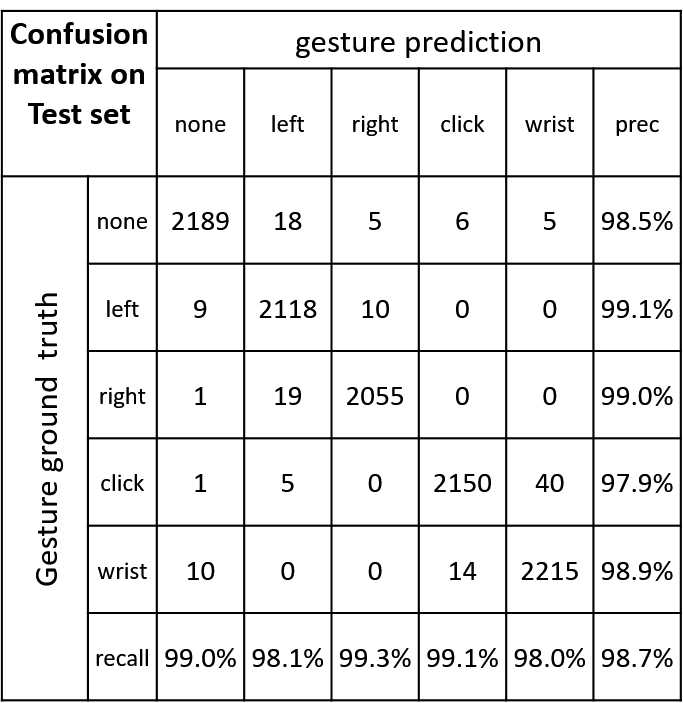}
  \caption{Confusion matrix on test set}
  \Description{Confusion Matrix on test set.}
  \label{ConfusionMatrix}
\end{figure}

In Figure \ref{ConfusionMatrix}, our model achieves 98\% average accuracy on test set.

\subsection{Error Analysis}
\begin{itemize}
\item LEFT is misclassified as RIGHT, and RIGHT is misclassified as LEFT. 
The gesture movement consists of three temporally overlapping phases: 
preparation, nucleus and retraction. 
The retraction phase of LEFT is a RIGHT, and the retraction phase of RIGHT is a LEFT. 
A better gesture segmentation preprocessing may help to reduce this kind of error.

\item CLICK is misclassified as LEFT and WRIST is misclassified as RIGHT. 
Recall Figure 4, the biggest difference between LEFT/RIGHT and CLICK/WRIST is the angle difference, 
when CLICK/WRIST gesture has large angle changes, they are easily misclassified as LEFT/RIGHT. 
More accurate angle estimation may help reduce this error.
\end{itemize}

\section{Conclusion}
FMCW radar is a low cost/high spatial/speed resolution sensors, 
it can detect anonymous object movements, 
suitable for privacy-concerned interaction application.
We designed a 20-layer residual network to recognize gestures, 
and the model could achieve 96\% accuracy on real-time test. 
In the future, we plan to support user defined gestures.

\bibliographystyle{ACM-Reference-Format}
\bibliography{references}

\end{document}